\documentstyle[12pt]{article}
    \def\l{\lambda}
    \def\m{\mu}
    \def\a{\alpha}

\begin{document}
\baselineskip=22pt plus 1pt minus 1pt

\begin{center}
{\Large \bf  Ground--$\gamma$ band mixing and odd--even staggering
in heavy deformed nuclei}
\end{center}
\medskip

\begin{center}
{\large N. Minkov$^*$\footnote[1]{e-mail: nminkov@inrne.bas.bg},
S. B. Drenska$^*$\footnote[2]{e-mail: sdren@inrne.bas.bg},
P. P. Raychev$^{*}$\footnote[3]{e-mail: raychev@bgcict.acad.bg},
R. P. Roussev$^*$\footnote[4]{e-mail: rousev@inrne.bas.bg}
and Dennis Bonatsos$^\dagger$\footnote[5]{e-mail:
bonat@mail.demokritos.gr}\\
\medskip
$^*$  Institute for Nuclear Research and Nuclear Energy, \\
72 Tzarigrad Road, 1784 Sofia, Bulgaria\\
\medskip
$^\dagger$ Institute of Nuclear Physics, N.C.S.R. ``Demokritos'',\\
GR-15310 Aghia Paraskevi, Attiki, Greece}
\end{center}
\bigskip\bigskip

\begin{abstract}
It is proposed that the odd-even staggering (OES) in the $\gamma$-
bands of heavy deformed nuclei can be reasonably characterized by a
discrete approximation of the fourth derivative of the odd-even
energy difference as a function of angular momentum $L$. This
quantity exhibits a well developed staggering pattern (zigzagging
behavior with alternating signs) in rare earth nuclei and actinides
with long $\gamma$- bands ($L\geq 10$). It is shown that the OES can
be interpreted reasonably as the result of the interaction of the
$\gamma$ band with the ground band in the framework of a Vector Boson
Model with SU(3) dynamical symmetry. The model energy expression
reproduces successfully the staggering pattern in all considered
nuclei up to $L=12-13$. The general behavior of the OES effect in
rotational regions is studied in terms of the ground--$\gamma$
band-mixing interaction, showing that strong OES effect occurs in
regions with strong ground--$\gamma$ band-mixing interaction. The
approach used allows a detailed comparison of the OES in $\gamma$
bands with the other kinds of staggering effects in nuclei and
diatomic molecules.
\end{abstract}
\bigskip\bigskip


\newpage

\section{Introduction}

Various types of deviation of nuclear collective rotations from the
well studied in first approximation pure rotational motion are known
\cite{BM75,RS80}.  They cause some higher order effects in the
structure of nuclear rotational spectra, such as the squeezing,
backbending and staggering.  The staggering effects represent
bifurcations of rotational bands into sequences of states differing
by several units of angular momentum.  Such effects are the odd--even
staggering observed in the collective $\gamma$- bands \cite{BM75},
the $\Delta L=1$, $\Delta L=2$ and $\Delta L=4$ staggering in
superdeformed nuclear bands \cite{Fli,Ced,Stag,WZ97} and the $\Delta
L=2$ staggering in the ground-state bands of normally deformed nuclei
\cite{WT97}. Similar staggering effects have been identified recently
in the rotational bands of diatomic molecules \cite{Stag,RDM97}.

In particular the odd--even staggering (OES) effect represents the
relative displacement of the odd angular momentum levels of the
$\gamma$- band with respect to their neighboring levels with even
angular momentum. This is a long known effect which is clearly
established in even--even nuclei \cite{BM75}. It, therefore, allows
one to test various collective models \cite{Bevod}. On the other
hand, the model interpretation of OES in the $\gamma$- bands could be
of use for the understanding of staggering effects in rotational
spectra as a whole.

In some studies the OES has been interpreted as a result of the
interaction between the even levels of the $\gamma$- band and
corresponding levels of a $\beta$- band \cite{Bevod}--\cite{Rie87}.
This consideration has been addressed to the SU(3) limit of the
Interacting Boson Model (IBM), in which the lowest $\beta$- and
$\gamma$- rotational bands interact in the framework of the same
irreducible representation (irrep), ($\l ,\m =2$), of the group SU(3)
\cite{IA}.

It is known that this approach comes to several complications
\cite{Bevod}:

i) In some nuclei the $\beta$- band, which should be responsible for
the OES in the $\gamma$- band is not observed experimentally
\cite{Rie87}, or in other cases it is not long enough
\cite{Sakai,Sood}.

ii) The model description of OES is limited in dependence on whether
the $\gamma$- band lies above the $\beta$-  band or not.

iii) The model analyses suggest increasing staggering in the
$\gamma$- band with increasing separation between  the $\gamma$- and
$\beta$- bands, which is not expected if staggering is due to the
interaction between these two bands.

Some of the above items could be dealt with in the SU(3) limit of the
IBM by using (for example) the four-body symmetry conserving
interactions introduced in \cite{BMI85} as well as by using in an
appropriate way the higher-order interactions introduced recently in
\cite{PVI99}.

On the other hand, it would be natural to extend the OES
investigation beyond the IBM classification scheme.  In this respect
the Vector Boson Model (VBM) with SU(3) dynamical symmetry
\cite{p:descr,a:over,p:matr} would be appropriate.

As an essential distinct from IBM the Vector Boson Model
classification scheme unites the $\gamma$- band together with the
ground band into the same split SU(3) multiplet. The similarities and
the differences between these two models as well as their mutual
complementary in different regions of rotational nuclei have been
outlined in Ref. \cite{MDRRB97}. An important feature of the Vector
Boson Model (VBM) scheme is that it provides a relevant way to study
the interaction between the ground and the $\gamma$- band
\cite{MDRRB98}.

It is, therefore, reasonable to check in the VBM framework whether
the OES in the $\gamma$- band could be interpreted as the result of
the ground--$\gamma$ band mixing interaction. Such an approach is
strongly motivated by the circumstance that items i) and ii) above
will be automatically removed. Indeed in  most of rotational nuclei
the even angular momentum levels of the $\gamma$- band have their
counterparts in the ground- band.  Also, the $\gamma$- band levels are
always placed above the corresponding levels of the ground- band, which
allows an equal treatment of the OES effect in all deformed
even--even nuclei. Moreover, recent investigations of the SU(3)
dynamical symmetry in deformed nuclei show a systematic  behavior of
the ground--$\gamma$ band interaction in dependence on the observed
SU(3) splitting, i.e. on the mutual disposition of the two bands
\cite{MDRRB97,MDRRB98}. On this basis one could study the possible
dependence of the OES effect on the ground--$\gamma$ band separation.
Such an analysis would be of use for the elucidating of item iii).

In the present work we respond to the above consideration. Our
purpose is to study the OES in the $\gamma$- bands of rotational
nuclei in terms of the ground--$\gamma$ band mixing interaction.  We
apply the Vector Boson Model (VBM) formalism in order to describe
this effect as well as to analyze its general behavior in rotational
regions.  Below it will be seen that in the ground--$\gamma$ band
coupling scheme of the VBM the OES effect appears in a rather natural
way. It will be shown that in this approach the staggering effect
exhibits a reasonable dependence on the ground--$\gamma$ band
splitting.  In addition, the study gives an insight into the recently
suggested \cite{MDRRB97,MDRRB98} transition between the
ground--$\gamma$ band coupling scheme of the VBM and the
$\beta$--$\gamma$ scheme of the IBM.

We remark that traditionally the OES is considered in terms of a plot
of the moment-of-inertia parameter versus the angular momentum $L$
\cite{Fi82,Fi84}.  On the other hand, the $\Delta L=2$ staggering
effects in nuclei and molecules have been established by introducing
a new relevant characteristic of rotational spectra, which is the
discrete approximation of the fourth derivative of the energy
difference between two levels with $\Delta L=2$ as a function of L.
In the present work we suggest that an analogous characteristic,
corresponding to $\Delta L=1$, will be appropriate for the case of
OES in the $\gamma$-bands.  As it will be seen below, this is a
rather convenient way to provide the analyses of the OES ($\Delta
L=1$) effect in deformed even--even nuclei.

In Sec. 2 the $\Delta L=1$ discrete derivation is introduced as an
OES effect characteristic. In Sec. 3 the Vector Boson Model (VBM) is
briefly presented and a relevant model expression for the staggering
quantity is obtained. The respective experimental and theoretical
staggering patterns for rare earth nuclei and actinides are presented
in Sec. 4. The results obtained are discussed in Sec. 5 and some
concluding remarks are given in Sec. 6.

\section{The odd-even staggering in a form of $\Delta L=1$ discrete
derivation}

In nuclear physics the transition energies between levels differing
by one or several units of angular momentum are experimentally well
determined quantities. In particular the transition energy:
\begin{equation}
\Delta E(L)=E(L+1)-E(L)\ ,
\label{tren}
\end{equation}
with $\Delta L=1$, carries essential information about the structure
of various nuclear collective bands  such as the $\gamma$-bands and
some negative parity bands of heavy  deformed nuclei. Its deviation
from the rigid rotor behavior can be measured by the quantity:
\begin{equation}
Stg(L)= 6\Delta E(L)-4\Delta E(L-1)-4\Delta E(L+1)+
\Delta E(L+2)+\Delta E(L-2).
\label{stag}
\end{equation}
In the case of a rigid rotor one can easily see that $Stg(L)$
is equal to zero. Moreover the terms of the second power in $L(L+1)$
also give zero in Eq.~(\ref{stag}). This is due to the fact that
$Stg(L)$ is the discrete approximation of the fourth
derivative of the function $\Delta E(L)$, i.e. the fifth derivative
of the energy $E(L)$. Therefore, any non-zero values of the quantity
$Stg(L)$ will indicate the presence of order higher than
$(L(L+1))^2$ from the regular rotational motion of the nuclear
system.

The above expression is introduced by analogy with the case of the
$\Delta L=2$ staggering in superdeformed nuclei \cite{Fli,Ced} and
$\Delta L=2,4$ staggering in diatomic molecules \cite{Stag}. The
respective quantities have been used properly in various theoretical
proposals for the explanation of these effects
\cite{SZG,MQ,Mag,Kota,Liu,Pav,Wu}, some of them
\cite{HM,Macc,PavFli,Doenau,Luo,Magi} using symmetry arguments which
could be of applicability to other physical systems as well.

On the above basis it is natural to apply the quantity
$Stg(L)$ to study the OES in the $\gamma$-bands of heavy
deformed nuclei, i.e. to interpret this effect in the form of $\Delta
L=1$ discrete derivation. Such an approach could be very useful in
providing a unified analysis of the different kinds of staggering
effects as well as in comparing their physical explanations.

Since for the $\gamma$-bands the experimental energy values are
known, Eq.~(\ref{stag}) can be written in the form:
\begin{eqnarray}
Stg(L) &=& 10E(L+1)+5E(L-1)+E(L+3) \nonumber \\
   &-& \left [10E(L)+5E(L+2)+E(L-2)\right ] \ ,
\label{stagen}
\end{eqnarray}
i. e. the quantity  $Stg(L)$ is simply determined by six band
level energies (with $L-2$, $L-1$, $\dots$, $L+3$). (Note that an
analogous  expression would not be useful for the staggering effects
in superdeformed nuclei and diatomic molecules where only the
transition energies are experimentally measured.)

As it will be seen in Sec. 4 Eq.~(\ref{stagen}) provides
a well developed staggering pattern (zigzagging behavior of
the function $Stg(L)$) for the experimentally observed $\gamma$
bands in the rare earth region and in the actinides.

\section{Odd-even staggering in the VBM}

Theoretically, the structure of the $\gamma$-bands of deformed nuclei
is well reproduced in the framework of the Vector Boson Model (VBM)
with SU(3) dynamical symmetry \cite{p:descr,a:over,p:matr,MDRRB97}.

The VBM is founded on the assumption that the low-lying collective
states of deformed even--even nuclei can be described with the use of
two distinct kinds of vector bosons, whose creation operators
$\mbox{\boldmath $\xi^{+}$}$ and $\mbox{\boldmath $\eta^{+}$}$ are
O(3) vectors and in addition transform according to two independent
SU(3) irreps of the type $({\l},{\m})=(1,0)$. The vector bosons
provide relevant constructions of the SU(3) angular momentum and
quadrupole operators like the bosons in the Schwinger realization of
SU(2) \cite{BieLou}. Also, they can be interpreted as quanta of
elementary collective excitations of the nucleus \cite{p:matr}.

The VBM Hamiltonian is constructed as a linear combination of three
basic O(3) scalars from the enveloping algebra of SU(3):
\begin{equation}
V=g_{1}L^{2}+g_{2}L\cdot Q\cdot L +g_{3}A^{+}A\ ,
\label{eq:v}
\end{equation}
where $g_{1}$, $g_{2}$ and $g_{3}$ are free model parameters; $L$ and
$Q$ are the angular momentum and quadrupole operators respectively;
and $A^{+}=\mbox{\boldmath $\xi^{+}$}^{2}\mbox{\boldmath
$\eta^{+}$}^{2}- (\mbox{\boldmath $\xi^{+}$}\cdot\mbox{\boldmath
$\eta^{+}$})^{2}$.

The Hamiltonian  (\ref{eq:v}) includes high, third ($L\cdot
Q\cdot L$) and fourth ($A^{+}A$) order effective interactions
and reduces the SU(3) symmetry to the rotational group SO(3).
It incorporates in a reasonable way the most important collective
properties of heavy deformed nuclei determined by their angular
momenta and quadrupole moments.

The model basis states
\begin{equation}
\label{eq:bast}
\left|\begin{array}{c}({\l},{\m})\\{\a},L,M\end{array}
\right\rangle\ ,
\end{equation}
corresponding to the $SU(3)\supset O(3)$ group reduction, are
constructed by means of the vector--boson operators and are
known as the basis of Bargmann--Moshinsky \cite{bm:bas,m:bas}.  The
quantum number $\a$ distinguishes the various O(3), O(2) irreps,
$(L,M)$, appearing in a given SU(3) irrep $(\l ,\m)$ and labels the
different bands of the multiplet.

In the VBM the ground- and the lowest $\gamma$- band belong to one
and the same SU(3) multiplet, in which $\l$ and $\m$ are even and
$\l\geq\m$. These bands are labeled by two neighboring integer values
of the quantum number $\a$.  The so defined multiplet is split with
respect to $\a$.

It is important to remark that while in the SU(3) limit of the IBM
the irreducible representations (irreps) ($\l, \m$) are restricted by
the total number of bosons describing the specific nucleus, in the
Vector Boson Model (VBM) the possible   ($\l ,\m$)- irreps of SU(3)
are {\em not restricted} by the underlying theory. On the other hand,
recently it has been shown \cite{MDRRB97} that some favored regions
of ($\l ,\m$) multiplets could be outlined through the numerical
analysis of the experimental data available for the ground and the
$\gamma$- collective bands of even--even deformed nuclei.  For the
favored irreps the VBM scheme gives a good description of the energy
levels and of the B(E2) transition ratios within and between the
bands. It should be emphasized that in the VBM the other collective
bands, in particular the lowest $\beta$-band, do not belong to the
same SU(3) irrep.

We remark that in the rare earth region and in the actinides the best
model descriptions correspond to (favored) SU(3) multiplets with ($\l
,\m =2$).

It is therefore reasonable to try to reproduce the fine
characteristics of the ground--$\gamma$ band mixing interaction in these
nuclei for $\mu =2$. That is why  the ($\l ,\m =2$) multiplets can be
naturally applied for the description and the interpretation of the
OES effect in the framework of the VBM.

In this case the model formalism  allows one to obtain simple
analytic expressions for the ground- and the $\gamma$- band energy
levels. For the  $(\l ,2)$- irreps the ground- and the $\gamma$-bands
are the only possible ones appearing in the corresponding SU(3)
multiplets. For the even angular momentum states the  Hamiltonian
matrix is always two-dimensional, while for the odd states of the
$\gamma$- band one has  single matrix elements.

The resolution of the standard eigenvalue equation gives the
following expressions for the even energy levels $E^{g}(L)$ and
$E^{\gamma}(L_{\mbox{even}})$ of the ground and the $\gamma$-band
respectively:
\begin{eqnarray}
E^{g}(L)&=& B+(A-BC)L(L+1)-
|B|\sqrt{1+aL(L+1)+bL^{2}(L+1)^{2}} \ ,
\label{Eground}  \\
E^{\gamma}(L_{\mbox{even}})&=& B+(A-BC)L(L+1)+
|B|\sqrt{1+aL(L+1)+bL^{2}(L+1)^{2}} \ .
\label{Egammae}
\end{eqnarray}
The odd $\gamma$ band levels, $E^{\gamma}(L_{\mbox{odd}})$ are
obtained in the form:
\begin{equation}
E^{\gamma}(L_{\mbox{odd}})=2B+AL(L+1) \ ,
\label{Egammao}
\end{equation}
where
    \begin{eqnarray}
    A &=& g_{1}-(2\l +5)g_{2}, \label{termA}\\
    B &=& 6(2\l +5)g_{2}-2(\l +3)^{2} g_{3},  \label{termB}\\
    C &=& \frac{g_3}{B} \ ,
    \label{termC}
    \end{eqnarray}
and
    \begin{eqnarray}
    a &=& -\frac{4}{B^{2}}\left\{(\l+3)[(\l+3)g_{3}-6g_{2}]g_{3}
      - 3(g_{3}-6g_{2})g_{2}\right\}, \label{trma} \\
    b &=& \frac{1}{B^{2}}\left(g_{3}-6g_{2}\right)^{2} \ ,
    \label{trmb}
    \end{eqnarray}
with $g_{1}$, $g_{2}$ and $g_{3}$ being the parameters of the
effective interaction (\ref{eq:v}). We remark that the application of
the VBM in rare earth nuclei and actinides \cite{MDRRB97} provides
$B>0$ in Eq.~(\ref{termB}) so that $|B|=B$ .

Further we rewrite Eqs. (\ref{Eground}), (\ref{Egammae}) and
(\ref{Egammao}) consistently, so as to obtain a unified expression
for all the $\gamma$- band energy levels:
\begin{eqnarray}
E^{g}&=&AL(L+1)\nonumber \\
     &-&B\left[\sqrt{1+aL(L+1)+bL^{2}(L+1)^{2}}+CL(L+1)-1\right] \ ,
\label{Egsb1}\\
E^{\gamma}&=&2B+AL(L+1) \nonumber \\
     &+&B\left[\sqrt{1+aL(L+1)+bL^{2}(L+1)^{2}}-CL(L+1)-1\right]
\left(\frac{1+(-1)^L}{2}\right) \ . \label{Egamma1}
\end{eqnarray}

So, $E^{\gamma}$ can be simply written
\begin{equation}
E^{\gamma}=2B+AL(L+1)+\frac{1}{2}BR(L)\left [1+(-1)^L\right ] \ ,
\label{Egamma}
\end{equation}
where
\begin{equation}
R(L)=\sqrt{1+aL(L+1)+bL^{2}(L+1)^{2}}-CL(L+1)-1 \ .
\label{termR}
\end{equation}

The last factor in Eq.~ (\ref{Egamma}) switches over $E^{\gamma}$
between the odd and the even states of $\gamma$- band. In such a way
the VBM gives a natural possibility to reproduce the parity effects
in the $\gamma$-band structure. Here, it is important to remark that
such a result is a direct consequence of the SU(3) dynamical symmetry
mechanism.

Now we are able to apply the above model formalism to reproduce
theoretically the $\Delta L =1$ discrete derivatives (\ref{stag}) and
(\ref{stagen}). After introducing Eq.~(\ref{Egamma}) into
Eq.~(\ref{stagen}) of the previous section, we obtain the following
model expression for the function $Stg(L)$:
\begin{eqnarray}
Stg(L)&=&
\frac{B}{2}\left( 10R(L+1)+5R(L-1)+R(L+3)\right) [1+(-1)^{L+1}]
\nonumber \\
&-&\frac{B}{2}\left( 10R(L)+5R(L+2)+R(L-2)\right)
[1+(-1)^{L}]  \ .
\label{VBMstag}
\end{eqnarray}

One can easily verify that the right-hand side of (\ref{VBMstag}) has
alternative signs as a function of the angular momentum values
$L=2,3,4,...$, i.e. it gives a regular model staggering pattern. In
addition, the amplitude of the staggering increases monotonously with
$L$. The signs and the amplitude are determined by the terms $BR(x)$,
($x=L-2,L-1,\dots ,L+3$) which depend on the high order effective
interactions in (\ref{eq:v}).

Eq.~(\ref{VBMstag}) allows one to study the OES effect in terms of
the VBM. Moreover the obtained result provides a reasonable
theoretical tool to interpret the OES effect in the meaning of a
$\Delta L =1$ staggering effect. In addition one could estimate
analytically the behavior of this effect in dependence on the nuclear
collective characteristics.

\section{The OES effect -- experimental pattern and theoretical
description}

We have applied Eq.~(\ref{stagen}) to the rare earth nuclei and to
the actinides for which the experimentally measured $\gamma$-bands
are long enough ($L\geq 10$), $^{156}$Gd \cite{Sood}, $^{156,160}$Dy
\cite{Sood}, $^{162}$Dy \cite{Sakai}, $^{162-166}$Er \cite{Sood},
$^{170}$Yb \cite{Arc98}, $^{228}$Th \cite{Web98}, $^{232}$Th
\cite{Sakai}. In all cases we obtain a clearly pronounced staggering
pattern, i.e. a zigzagging behavior of the quantity $Stg(L)$ as a
function of angular momentum. This is shown in Figs. 1 -- 10 (lines
with squares). Generally, one observes a regular change in the signs
of $Stg(L)$ between the odd and even levels with the amplitude
increasing with angular momentum $L$ up to $L=12-13$.

In such a way the OES effect in the $\gamma$- bands appears in the
form of a $\Delta L=1$ staggering, which is consistent with the
consideration of other staggering effects in nuclei and diatomic
molecules. We remark that for all $\gamma$- bands under study the
experimental uncertainties in the pattern (\ref{stagen}) are
negligible.

We remark that for the different nuclei the staggering amplitude
varies in a rather wide range. For example, for $L=8$ the quantity
$Stg(8)$ obtains the smallest absolute value $0.013$ MeV (for
$^{166}$Er, Fig. 7) and the largest value $0.467$ MeV (for
$^{156}$Gd, Fig. 1).

Interesting staggering patterns are observed in the nuclei $^{164}$Er
(Fig. 6) and $^{170}$Yb (Fig. 8) for which the $\gamma$ bands are
longest, $L_{\mbox{max}}=19$ for $^{164}$Er \cite{Sood},
$L_{\mbox{max}}=17$ for $^{170}$Yb \cite{Arc98}. In these cases the
staggering amplitude initially increases as a function of angular
momentum up to $L=8-10$ and then begins to decrease. Further, at
$L=14$, an irregularity in the alternative signs of the quantity
$Stg(L)$ occurs.

We have applied the model expression  Eq.~(\ref{VBMstag}) to describe
theoretically the staggering pattern of all the nuclei under
consideration. For this purpose we use the sets of model parameters
$g_{1}$, $g_{2}$, $g_{3}$ and $\lambda$, obtained after fitting Eqs.
(\ref{Egsb1}) and (\ref{Egamma1}) to the experimental ground- and
$\gamma$- band levels up to $L=10-12$. The values of these parameters
are listed in Table 1 together with the corresponding rms deviations
measured by
\begin{equation}
\label{eq:se}
\sigma_{E} =\sqrt{\frac{1}{n_{E}}\sum_{L,\nu }
\left( E_{\nu}^{Th}(L)-E_{\nu}^{Exp}(L)\right) ^{2}},
\end{equation}
where $n_{E}=n_{g}+n_{\gamma}$ is the total number of the levels used
in the fit and $\nu =g,\gamma$ labeling the ground and the
$\gamma$-band levels respectively.

For all the nuclei the theoretical pattern reproduces the alternating
signs as well as the general increase in the staggering amplitude as
a function of $L$ up to $L=12-13$.

Good description of the staggering pattern is obtained for the
nuclei $^{156,160,162}$Dy (Figs. 2--4), $^{166}$Er (Fig. 7),
$^{228}$Th (Fig. 9), $^{232}$Th (Fig. 10). Also for the nuclei
$^{164}$Er (Fig. 6) and $^{170}$Yb (Fig. 8) the staggering pattern is
well reproduced up to $L=13$ (Fig. 6 and Fig. 8 respectively), i.e.
up to the appearance of the sign irregularity.

For two of the nuclei, $^{156}$Gd and $^{162}$Er, the difference
between the theoretical and the experimental staggering magnitude
is noticeably larger than for the other nuclei under study (Fig.
1 and Fig. 5). In the next section it will be shown, that such
disagreement could be referred to the rather strongly perturbed
rotational structure of the respective ground and  $\gamma$-
bands.  As it is seen from Table 1, this circumstance reflects on
the quality of the model energy descriptions obtained for both
nuclei. The respective RMS factors are relatively larger
($\sigma_{E}=40$~keV for $^{156}$Gd and $\sigma_{E}=32.7$~keV for
$^{162}$Er) than the ones in the other nuclei.

On the other hand we remark, that for the obtained sets of
parameters (Table 1) our fitting procedures guarantee correct
reproduction of all the B(E2) transition rates available for the
ground- and the $\gamma$-bands. This is extensively demonstrated in
our previous paper \cite{MDRRB97} where the advantages of the Vector
Boson Model description (compared to other collective models) are
pointed out. That is why we do not concentrate on such considerations
in the present study. In Table 2, as an illustration,  we show the
theoretical ground- and $\gamma$-band energy levels and the attendant
B(E2) transition ratios of the nucleus $^{166}$Er which correspond to
the parameter set given in Table 1. The following B(E2) transition
ratios are included in our numerical procedures \cite{MDRRB97}:
    \begin{eqnarray}
    \label{eq:rat}
    R_{1}(L)&=&\frac{B(E2;L_{\gamma}\rightarrow L_{g})}
    {B(E2;L_{\gamma}\rightarrow (L-2)_{g})} \ , \ \mbox{L even,}\\
    R_{2}(L)&=&\frac{B(E2;L_{\gamma}\rightarrow (L+2)_{g})}
    {B(E2;L_{\gamma}\rightarrow L_{g})} \ , \ \mbox{L even,}\\
    R_{3}(L)&=&\frac{B(E2;L_{\gamma}\rightarrow (L+1)_{g})}
    {B(E2;L_{\gamma}\rightarrow (L-1)_{g})} \ , \ \mbox{L odd,}\\
    \label{eq:intra}
    R_{4}(L)&=&\frac{B(E2;L_{g}\rightarrow (L-2)_{g})}
    {B(E2;(L-2)_{g}\rightarrow (L-4)_{g})}\ ,  \ \mbox{L even,}
    \end{eqnarray}
where the indices $g$ and $\gamma$ label the ground- and the
$\gamma$-band levels respectively.

As it is seen from Table 2, very good agreement between the
theoretical and experimental data is obtained. This example
demonstrates that for the same set of model parameters the staggering
effect is reproduced in consistency with the other electro-magnetic
and energy characteristics of the ground- and the $\gamma$ band.

The results presented show that a relevant description of the
OES effect ($\Delta L=1$ staggering) in the $\gamma$-bands of rare
earth nuclei and actinides is possible in the framework of the VBM
with SU(3) dynamical symmetry.

\section{Discussion}

We are now able to analyze several important characteristics of the
fine rotational structure of the $\gamma$-bands together with the
respective nuclear collective properties hidden behind them. Although
the number of considered nuclei (10 nuclei) does not allow one to
provide any detailed systematics, our study leads to a consistent
theoretical interpretation of all the available experimental
information concerning the $\Delta L=1$ (OES) staggering effect in
$\gamma$- bands. (Note that two newest sets of data are used for
$^{170}$Yb \cite{Arc98} and $^{228}$Th \cite{Web98}).

The analysis of the experimental $\Delta L=1$ staggering amplitude
obtained in rare earth nuclei shows that it is generally larger for
the nuclei placed in the beginning of this rotational region compared
to the midshell nuclei. For example the staggering amplitude in
$^{156}$Gd is more than one order of magnitude larger than the ones
in $^{166}$Er (See Figs. 1 and 7). Also, a gradual decrease of the
amplitude towards the midshell region is observed for the three Er
isotopes, $^{162}$Er, $^{164}$Er and $^{166}$Er (with $Stg(8)=0.425$
MeV, $Stg(8)=0.251$ MeV and $Stg(8)=-0.013$ MeV respectively; see
Figs. 5, 6 and 7).

The above observation is consistent with the general behavior of the
nuclear rotational properties in the limits of the valence shells. It
is well known that towards the midshell region these properties are
better revealed so that any kind of deviations from the regular
rotational band--structures should be smaller. In this respect the
weaker $\Delta L=1$ staggering effect observed in the rare earth
midshell isotopes is quite natural.

On the other hand, such a behavior of the staggering effect can be
reasonably interpreted in terms of the ground--$\gamma$ band
interaction. It has been shown in the Vector Boson Model framework
that this interaction systematically decreases towards the  middle of
rotational regions \cite{MDRRB97,MDRRB98}. Thus the weaker mutual
perturbation of these two bands in the midshell region is consistent
with the respectively good rotational behavior of the $\gamma$-band.

Also, it has been established that the ground--$\gamma$ band mixing
interaction in heavy deformed nuclei is correlated with the energy
separation between the two bands. In the SU(3) dynamical symmetry
framework this separation corresponds to the splitting of the SU(3)
multiplet and is measured by the ratio:
\begin{equation}
\label{eq:splitL}
\Delta E_{L}=\frac{E_{L}^{\gamma}-E_{L}^{g}}{E_{2}^{g}}\ ,
\end{equation}
which characterizes the magnitude of the energy differences between
the even angular momentum states of the ground- and the $\gamma$- band.

For example the experimental $\Delta E_{2}$ ratios vary within the
limits $5\leq\Delta E_{2}\leq 20$, for the nuclei of rare earth
region, and $13\leq\Delta E_{2}\leq 25$, for the actinides.

It has been found that the  $\Delta E_{L}$ ratio generally increases
towards the middle of a given rotational region in consistency with
the decrease in the  ground--$\gamma$ band mixing interaction and
corresponds to an increase in the SU(3) quantum number $\l$.

On the above basis it has been established that for nuclei with a
weak SU(3) splitting ($\Delta E_{2} \leq 12$ for rare earth nuclei
and $\Delta E_{2} \leq 15$ for actinides) the ground and the $\gamma$
bands are strongly coupled in the framework of the SU(3) dynamical
symmetry, with $\l$ obtaining favored values in the region
$\l=14-20$.

Now we remark that the nuclei of the present study indeed belong
to this kind of SU(3) dynamical symmetry nuclei. In such a way the
relatively strong ground--$\gamma$ band interaction in these nuclei can
be considered as the reason causing the observed OES ($\Delta L=1$
staggering).

For the nuclei with strong SU(3) splitting ($\Delta E_{2} >12$ for
rare earth nuclei, and $\Delta E_{2} > 15$ for actinides) the ground and
the $\gamma$ bands are weakly coupled in the framework of the SU(3)
dynamical symmetry, with $\l >60$. Typical examples for this kind of
nuclei are $^{172}$Yb (with $\Delta E_{2} =17.6$) and $^{238}$U (with
$\Delta E_{2} =20.3$). It has been suggested that for these nuclei
the ground- and the $\gamma$- band could be separated in different SU(3)
irreps. Then the OES can be possibly interpreted as the interaction
of the $\gamma$ band with the $\beta$-band as it is done in the
framework of the Interacting Boson Model \cite{IA,Bevod}. We remark
that in these cases the $\gamma$- bands are not long enough in order
to study the OES in the form of $\Delta L=1$ staggering.

The above consideration is strongly consistent with the possibility
for a transition \cite{MDRRB97,MDRRB98} from the ground--$\gamma$
band coupling scheme of the present VBM, which is more appropriate
near the ends of the rotational regions, to the IBM classification
scheme \cite{IA} with $\beta$--$\gamma$ band coupling, which is more
relevant in the midshell nuclei. It clearly illustrates
that both model schemes are mutually complementary in the different
regions of rotational nuclei. Moreover, we remark that for the
nuclei considered in the present study the proposed ground--$\gamma$
band mixing interpretation of the OES effect is {\em unique}.

At this point it is reasonable to discuss the general behavior of the
OES effect in heavy deformed nuclei in terms of the SU(3) dynamical
symmetry characteristics.

Here we refer to the so called SU(3) contraction limits, in which the
algebra of SU(3) goes to the algebra of the semi-direct product
T$_{5}\wedge$SO(3), i.e. $SU(3)\rightarrow T_{5}\wedge SO(3)$ (T$_5$
is the group of 5-dimensional translations generated by the
components of the SU(3)- quadrupole operators)
\cite{Gilmore,CBVR86,CDL88,RVC89,Mukerjee,JPD93}. (Generally, the
contraction limit corresponds to a singular linear transformation of
the basis of a given Lie algebra. The transformed structure constants
approach well-defined limits and a new Lie algebra, called contracted
algebra, results \cite{Gilmore}. The original and the contracted
algebras are not isomorphic.)

In the VBM collective scheme the SU(3) contraction corresponds to the
following two limiting cases:

(i) $\l\rightarrow\infty $, with $\m$ finite;

(ii) $\l\rightarrow\infty $, $\m\rightarrow\infty $, with $\m\leq\l$,

It has been shown that in these limits the ground--$\gamma$ band mixing
gradually disappears so that the corresponding SU(3) multiplets are
disintegrated into distinct noninteracting bands \cite{MDRRB98}. It
follows that all fine spectroscopic effects based on the band-mixing
interactions, such as the OES effect, should be reduced towards the
SU(3) contraction limits.

Indeed, one can easily verify that in the limit (i) the OES effect
should be not observed. In fact, for this case we have deduced
analytically that the terms of the type $BR(L)$ which determine the
quantity $Stg(L)$ in Eq.~(\ref{VBMstag}) go to zero as
\begin{equation}
\lim_{\l\rightarrow\infty}BR(L)=\lim_{\l\rightarrow\infty}
\frac{3(g_{2}g_{3}-3g_{2}^{2})}{g_{3}\l^{2}}L(L+1)=0 \ .
\label{limst1}
\end{equation}
(This analytic limit is obtained by using the explicit expressions
(\ref{termB}) and (\ref{termR}) with (\ref{trma}), (\ref{trmb}) and
(\ref{termC}).) It follows that the staggering amplitude goes to zero
when $\l$ increases to infinity. This is illustrated in Fig. 11,
where the model staggering pattern is plotted for $\l =20$, $\l =40$
and $\l =60$ and fixed (overall) values of the model parameters
$g_{2}=-0.2$ and $g_{3}=-0.25$.

We remark that the first limiting case, $\l\rightarrow\infty $ with
$\m =2$, is physically reasonable for midshell nuclei, which are
characterized by large values of the quantum number $\l$ (with $\m
=2$) and strong ground--$\gamma$ band splitting.
However, we emphasize that for the nuclei (under study) where
odd-even staggering is observed the SU(3) scheme of the Vector Boson
Model works far from the SU(3) contraction limits.

In this way the present model approach completely resolves the
problem iii) stated in Sec. 1. Indeed, we obtain that the increase in
the separation between the ground and the $\gamma$ band (i.e. the
SU(3) splitting) is correlated with the respective decrease in the
magnitude of the OES effect. In addition, the lack of staggering in
the SU(3) contraction limit (where the SU(3) multiplets are
disintegrated) endorses our conclusion that for the nuclei under
study this effect should due to the SU(3) coupling of the $\gamma$
band together with the ground band.  In this respect the extents to
which the observed phenomenon could be interpreted as a result of
forced ground--$\gamma$ band mixing become clear.

On the other hand, the area of applicability of the present VBM
scheme is also clear. Besides the contraction limit, our model
interaction comes to another restriction which appears naturally
towards the transition region. This is indicated by the
circumstance that it suffers in the reproduction of the large
staggering amplitudes in the nuclei $^{156}$Gd and $^{162}$Er
($Stg(8)\sim 0.4-0.5$MeV, see Figs. 1 and 5) which are placed far
from the middle of the rotational region. (More precisely,
$^{156}$Gd is close to the beginning of the region, while
$^{162}$Er is in the beginning of the respective group of
rotational isotopes.) Actually, our analyses suggest that in
these cases the ground--$\gamma$ mixing is far stronger than a
small perturbation to the respective rotational band structures.
This is supported by the observation of large experimental
ground--$\gamma$ interband transition probabilities
\cite{MDRRB98}. In such a way, the above difficulty of our model
description could be referred to the standard problem of the
strong perturbation interactions.

Also, it is important to remark that the staggering
pattern could be influenced by the presence of $\beta$--$\gamma$
interaction. Although for the nuclei considered the latter should be
essentially weaker than the ground--$\gamma$ interaction, its
involvement in a more general study would be of interest. Since the
respective quantitative analysis oversteps the capacity of the
present vector boson scheme (which does not include the $\beta$-
rotational band) here we only give an idea about two future possible
approaches to the problem:

1) A symplectic Sp(6,R) extension of the present  VBM with
involvement of appropriate mixing between the different SU(3)
multiplets;

2) IBM analysis with consistent application of higher order
interactions in both the SU(3) limit (with the standard
$\beta$--$\gamma$ coupling) \cite{BMI85} and the recently proposed
O(6) scheme \cite{PVI99} with a cubic quadrupole interaction
$(\hat{Q}\times\hat{Q}\times\hat{Q})^{(0)}$. The latter seems to be
very promising for a relevant IBM treatment of the ground--$\gamma$
band interaction in deformed nuclei.

So, in these ways one could try to improve the theoretical
description of the observed staggering patterns. Of course, the price
to be paid is the more complicated model formalism and the larger
number of parameters. Also, one should not expect essential
improvement in the patterns with large experimental amplitudes
($Stg(8)\sim 0.4-0.5$~MeV) where the problems due (as has already
been mentioned) to the generally stronger perturbed
rotational structure of the $\gamma$- band.

Let us now consider the two long staggering patterns in $^{164}$Er
(Fig. 6) and $^{170}$Yb (Fig. 8) for which sign irregularities are
observed at $L=14$. It is known that at this angular momentum the
structure of the $\gamma$-band of $^{164}$Er is changed. This is
interpreted as the result of a crossing with another band known as a
super band \cite{Jon78,Yat80}. The same phenomenon is considered to
be responsible for the backbending effect observed in this $\gamma$-
band \cite{Bon85}. On the above basis our analysis suggests that a
similar situation is realized in the $^{170}$Yb  $\gamma$-band
\cite{Arc98}. In fact the backbending effect is beyond the scope of
the presently used VBM with SU(3) dynamical symmetry, which explains
the reason why our theoretical description is restricted up to $L\leq
12-13$. It is however remarkable that the experimentally determined
quantity $Stg(L)$ gives an excellent indication for the
presence of bandcrossing effects.

It is important to emphasize the meaning of the introduction of the
$\Delta L=1$ characteristics of the $\gamma$- bands. Indeed the
moment-of-inertia versus the angular momentum analyses reasonably
indicate the presence of the OES effect \cite{Fi82,Fi84}. On the
other hand, the use of the fourth derivative of the odd-even energy
differences gives a rather accurate quantitative measure to estimate
the magnitude of this effect for a given angular momentum or given
region of angular momenta. In such a way the role of the band-mixing
interactions could be correctly taken into account. Moreover, the
well determined staggering amplitudes together with the clearly
established alternating signs pattern allow one to provide various
quantitative analyses (as the present one) of the fine structure of
nuclear collective bands as a whole.

\section{Conclusion}

We have studied the OES effect in the $\gamma$- bands of even-even
deformed nuclei in terms of the discrete approximation of the fourth
derivative of the $\Delta L=1$ (odd-even) energy difference. The
staggering pattern obtained in several rare earth nuclei and
actinides is clearly pronounced (with the respective experimental
uncertainties being negligible) and can be referred to as $\Delta
L=1$ staggering. Its form is essentially similar to the one seen for
the other kinds of staggering observed in nuclei and diatomic
molecules. The most common feature of all staggering patterns is the
initial increase of the amplitude as a function of angular momentum
followed by its alternations with $L$ as well as by possible
occurrence of sign irregularities. Thus in all cases the staggering
pattern reflects the fine structure of rotational bands and gives a
rather natural indication for some singular changes such as the
bandcrossing effects.

We have shown that the OES can be interpreted reasonably as the
result of the interaction of the $\gamma$ band with the ground
band in the framework of the Vector Boson Model with SU(3) dynamical
symmetry. The model energy expression reproduces adequately the
staggering pattern in the considered nuclei below the backbending
region $L=12-13$. On the above basis we were able to study the
general behavior of the OES effect in rotational regions in terms of
the ground--$\gamma$ band-mixing interaction. As a result we have
established that the increase in the separation between the ground and
the $\gamma$ band towards the midshell region is correlated with the
respective decrease in the magnitude of the OES effect. Thus we
explained the presence of the well developed OES patterns in nuclei
with relatively weak SU(3) splitting (strong ground--$\gamma$ band
coupling) as well as the decrease in the staggering amplitude (even
the absence of the effect) for the strongly split SU(3) multiplets in
midshell nuclei.

The approach presented gives a rather general prescription for
analysis of various fine characteristics of rotational motion in
quantum mechanical systems. In this respect it allows a detailed
comparison of the different kinds of staggering effects in nuclei and
diatomic molecules. A future unified interpretation and/or treatment
of these fine effects could be possible on this basis.

\bigskip
{\Large{\bf Acknowledgments}}
\medskip

One of authors (NM) would like to thank to Prof. J. H\"{u}fner who
supported his participation on the International Conference ``50
Years of the Nuclear Shell Model'' (Heidelberg 1999) where several
useful discussions of the present paper have been realized. The
discussions with Prof. F. Iachello  as well as with Dr. J. Hirsch are
acknowledged with thanks. We also thank to Prof. K. Sugawara-Tanabe
for illuminating discussions. This work has been supported by the
Bulgarian National Fund for Scientific Research under contract no
MU--F--02/98.

\newpage

\begin{table}
\caption{The parameters of the fits of the energy levels of the
ground- and the $\gamma$- bands (Eqs. (\protect\ref{Egsb1}) and
(\protect\ref{Egamma1})) of the nuclei investigated are listed for
the $(\l ,2 )$ multiplets which provide the best model descriptions.
The Hamiltonian parameters $g_{1}$, $g_{2}$ and $g_{3}$ (Eq.
(\protect\ref{eq:v})) and the RMS quantities $\sigma_{E}$ (Eq.
(\protect\ref{eq:se})) are given in keV. The numbers of the ground-
and the $\gamma$- band levels used in the fit, $n_{g}$ and
$n_{\gamma}$ respectively, are also given.}
    \bigskip
    \begin{center}
    \begin{tabular}{ccccccccccc}
    \rule{0em}{2.2ex}
   & & & & & & \\
\hline\hline
$Nucl$&$\l $&$g_{1}$&$g_{2}$&$g_{3}$&$\sigma_{E}$&$n_{g}
$&$n_{\gamma}$ \\
\hline
${}^{156}\rm Gd
$&$16$&$7.179$&$-0.138$&$-0.819$&$40.0$&$ 5 $&$ 9 $  \\
${}^{156}\rm Dy
$&$14$&$2.241$&$-0.380$&$-0.875$&$47.8$&$ 5 $&$ 9 $  \\
${}^{160}\rm Dy
$&$16$&$7.176$&$-0.124$&$-0.682$&$24.5$&$ 5 $&$ 11$  \\
${}^{162}\rm Dy
$&$16$&$8.980$&$-0.059$&$-0.606$&$21.3$&$ 5 $&$ 11$  \\
${}^{162}\rm Er
$&$16$&$9.264$&$-0.116$&$-0.637$&$32.7$&$ 5 $&$ 11$  \\
${}^{164}\rm Er
$&$16$&$8.345$&$-0.117$&$-0.597$&$21.4$&$ 5 $&$ 11$  \\
${}^{166}\rm Er
$&$16$&$3.553$&$-0.210$&$-0.570$&$23.4$&$ 5 $&$ 11$  \\
${}^{170}\rm Yb
$&$18$&$5.824$&$-0.140$&$-0.663$&$26.6$&$ 5 $&$ 11$  \\
${}^{228}\rm Th
$&$20$&$3.169$&$-0.098$&$-0.470$&$13.6$&$ 5 $&$  9$  \\
${}^{232}\rm Th
$&$20$&$-7.564$&$-0.315$&$-0.438$&$22.2$&$9 $&$ 11$  \\
    \hline\hline
    \end{tabular}
    \end{center}
    \label{tab:favor}
    \end{table}

\ \ \ \ \ \
\newpage


\begin{table}
\caption{Theoretical and experimental energy levels and transition
ratios (Eqs. (\protect\ref{eq:rat}) -- (\protect\ref{eq:intra})) for
the nucleus $^{166}$Er, corresponding to the multiplet $(16,2)$ and
the set of parameters given in Table 1. The experimental data (used
in the fits) for the energy levels are taken from
\protect\cite{Sood}, while the data for the E2 transitions are from
\protect\cite{166Er1,166Er2,NDS166}. The numbers in brackets refer to
the uncertainties in the last digits of the experimental ratios.}
    \bigskip
    {\scriptsize
    \begin{center}
    \begin{tabular}{cc cccccccccccc}
& & & & & & & & & & & & &\\
\hline\hline
$L$&$E_{g}^{Th}$&$E_{g}^{Exp}$&$E_{\gamma}^{Th}$&$E_{\gamma}^{Exp}$&
$R_{1}^{Th}$&$R_{1}^{Exp}$&$R_{2}^{Th}$&$R_{2}^{Exp}$&
$R_{3}^{Th}$&$R_{3}^{Exp}$&$R_{4}^{Th}$&$R_{4}^{Exp}$ \\
\hline
2& 74.8 & 80.6 &   797.8 & 785.9 & 1.75 & 1.86(10) & 0.08 & 0.097(8)
&--&--&--&--& \\
3&--&--&           865.7  & 859.4 &--   &--       &--   &--
& 0.64 & 0.72(6) &--&-- \\
4& 249.2 & 265.0 & 956.4 & 956.2 & 4.8 & 5.72(47) & 0.18 & 0.26(7)
&--&--& 1.39 & 1.45(16)  &        \\
5&-- &--         & 1069.5& 1075.3 &--&--&--&--& 1.20 & 1.43(15)
&--&--&                \\
6& 523.1 & 545.4 & 1205.8& 1215.9 & 8.29 & 12.25(75) & 0.28& 0.28
&--&--& 1.05 & 1.12(22)   &        \\
7&--&--          & 1363.9 & 1376.0 &--     &--         &--
&--&1.89&--&--&--&                 \\
8& 896.2 & 911.2 & 1546.3 & 1555.7& 13.45 & 20.9(45)&0.42&--
&--&--   & 0.97 & 1.05(28)&        \\
9&--   &--       & 1748.9 & 1751.1&--      &--       &--      &--
&2.87    &--       &--      &--     & \\
10& 1368.0 & 1349.6 & 1978.3 & 1964.0&22.7 &--       &0.61    &--
&--    &--       &0.92      &1.02(27) & \\
\hline\hline
\end{tabular}
    \end{center}
 }
\end{table}

\ \ \ \ \ \

\newpage

    \begin{center}
    {\bf Figure Captions}
    \end{center}
    \bigskip\bigskip

\noindent
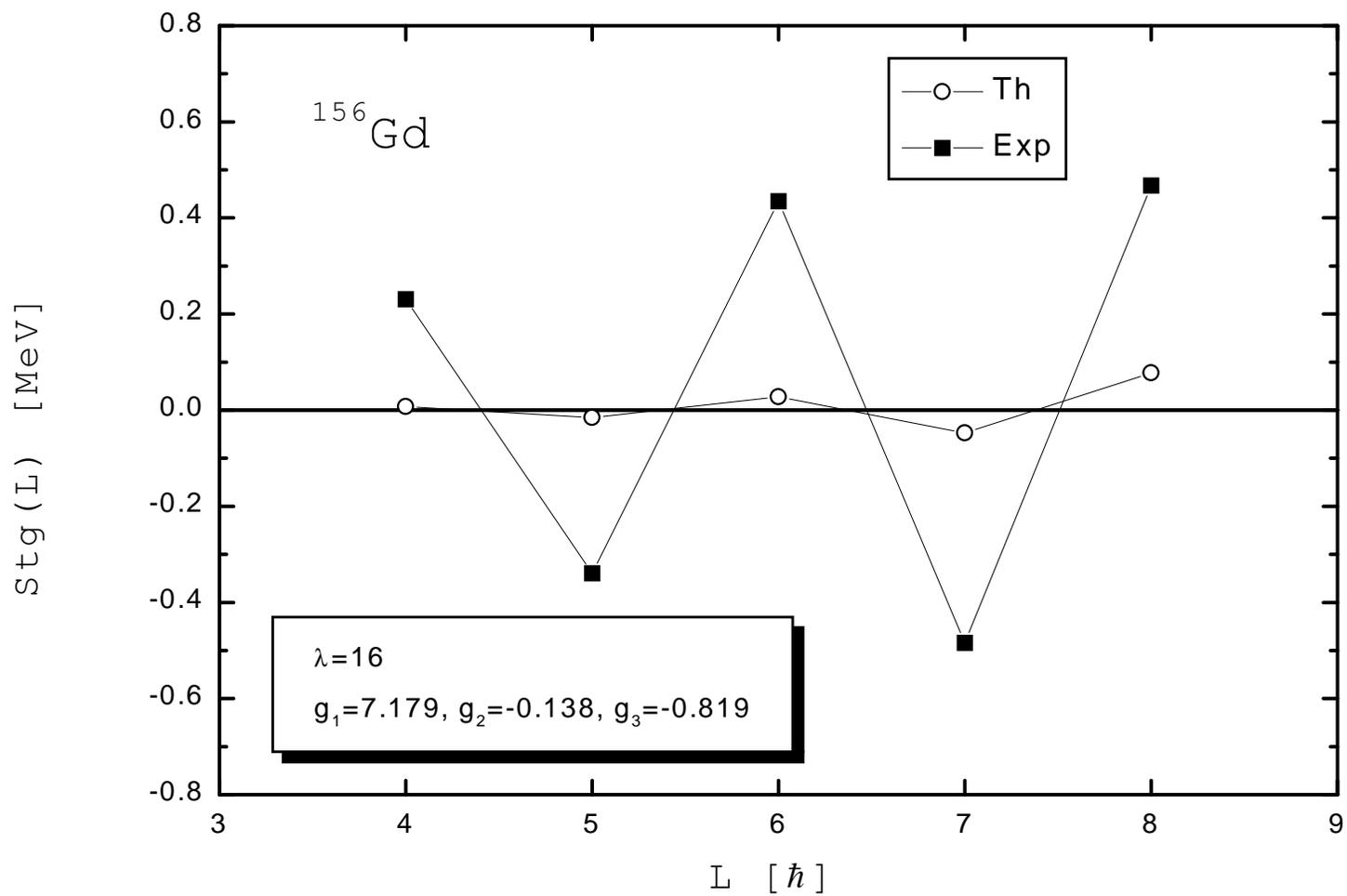
{\bf Figure 1.} The experimental and the theoretical values of the
quantity $Stg(L)$, Eq.~(\ref{stagen}) and Eq.~(\ref{VBMstag})
respectively, obtained for the $\gamma$- band of $^{156}$Gd are
plotted as functions of the angular momentum. The experimental data
are taken from \cite{Sood}.
\medskip

\noindent
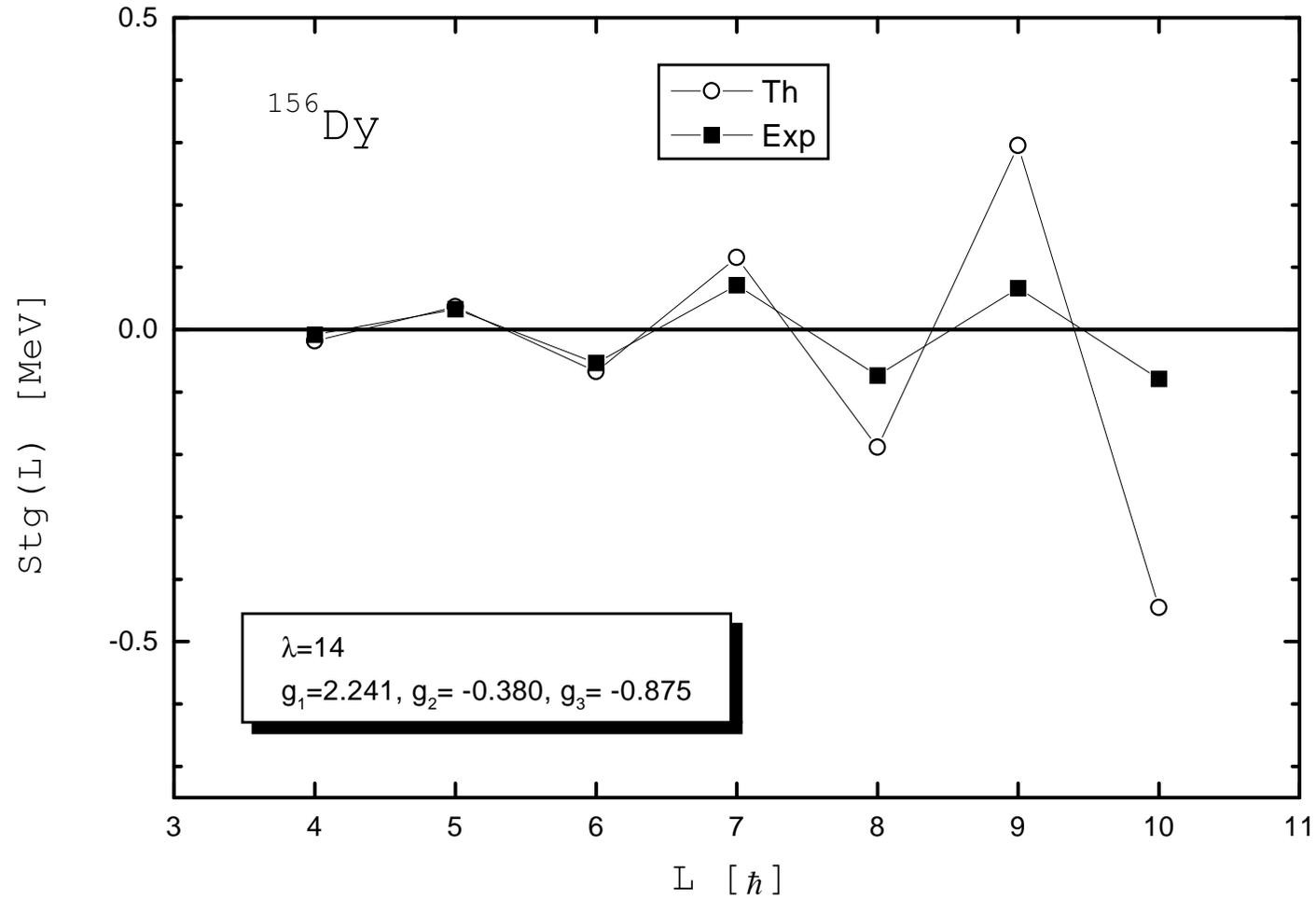
{\bf Figure 2.} The same as Fig. 1, but for the $\gamma$- band of
$^{156}$Dy.
\medskip

\noindent
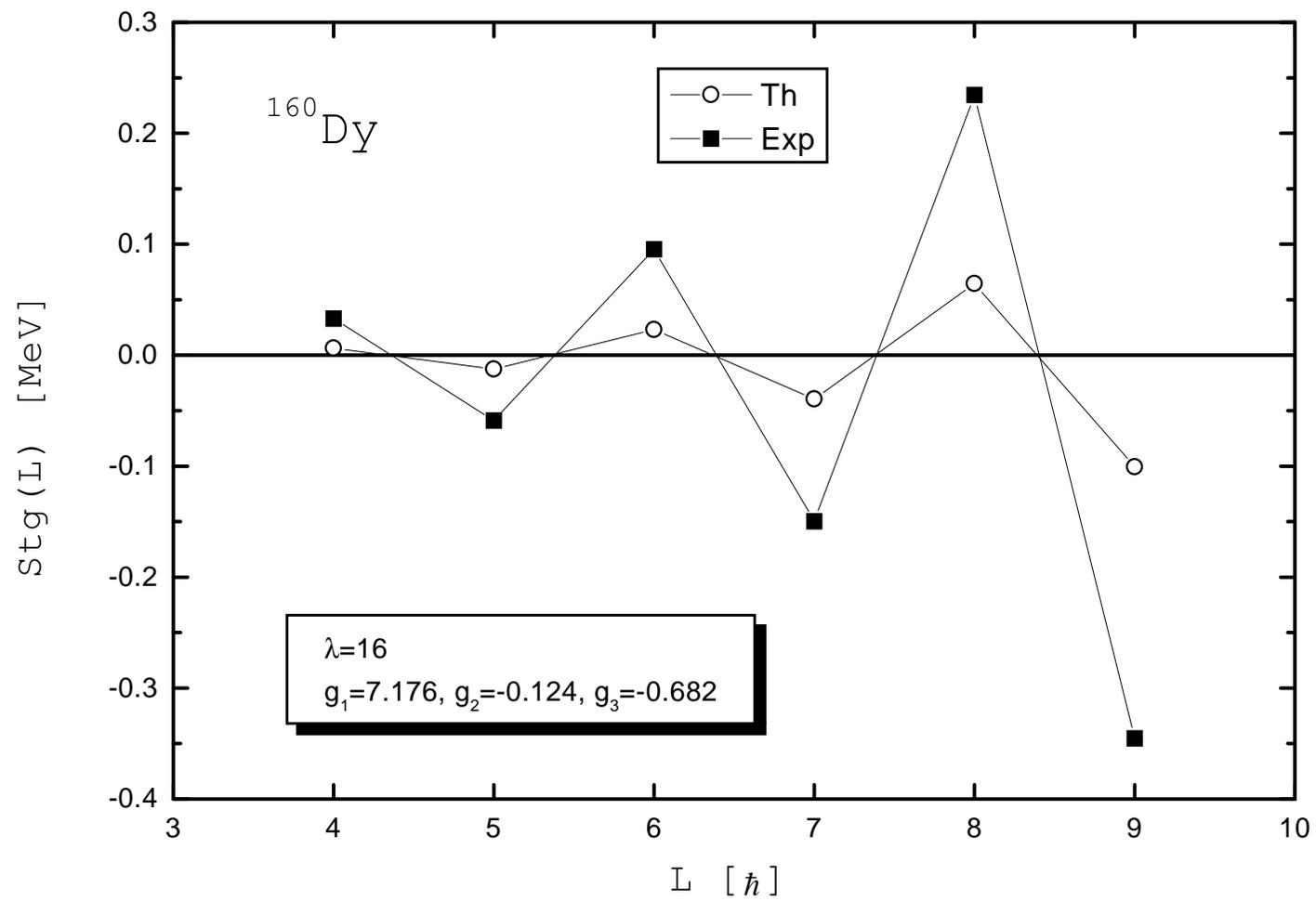
{\bf Figure 3.} The same as Fig. 1, but for $^{160}$Dy.
\medskip

\noindent
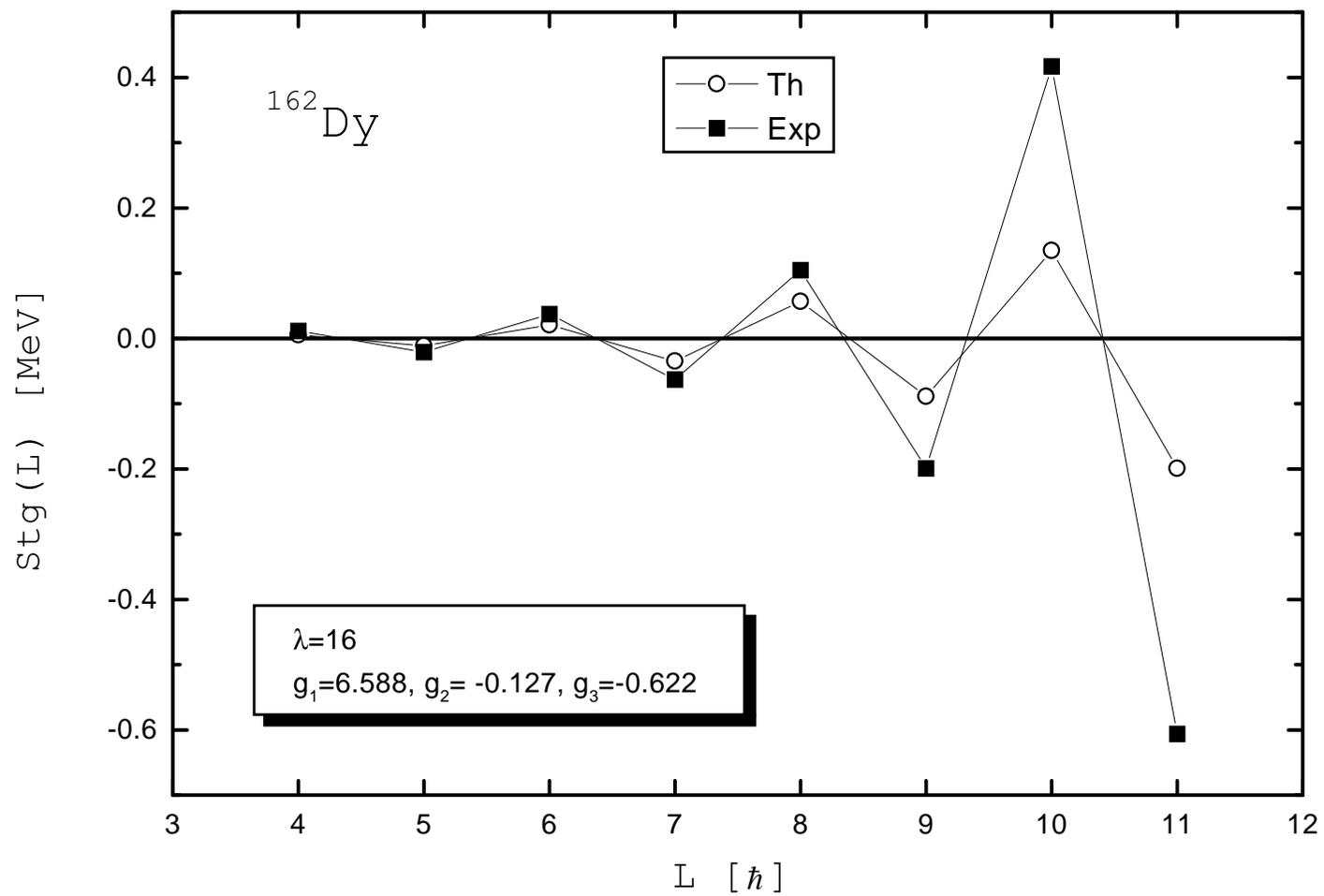
{\bf Figure 4.} The same as Fig. 1, but for $^{162}$Dy.
The experimental data are taken from \cite{Sakai}.
\medskip

\noindent
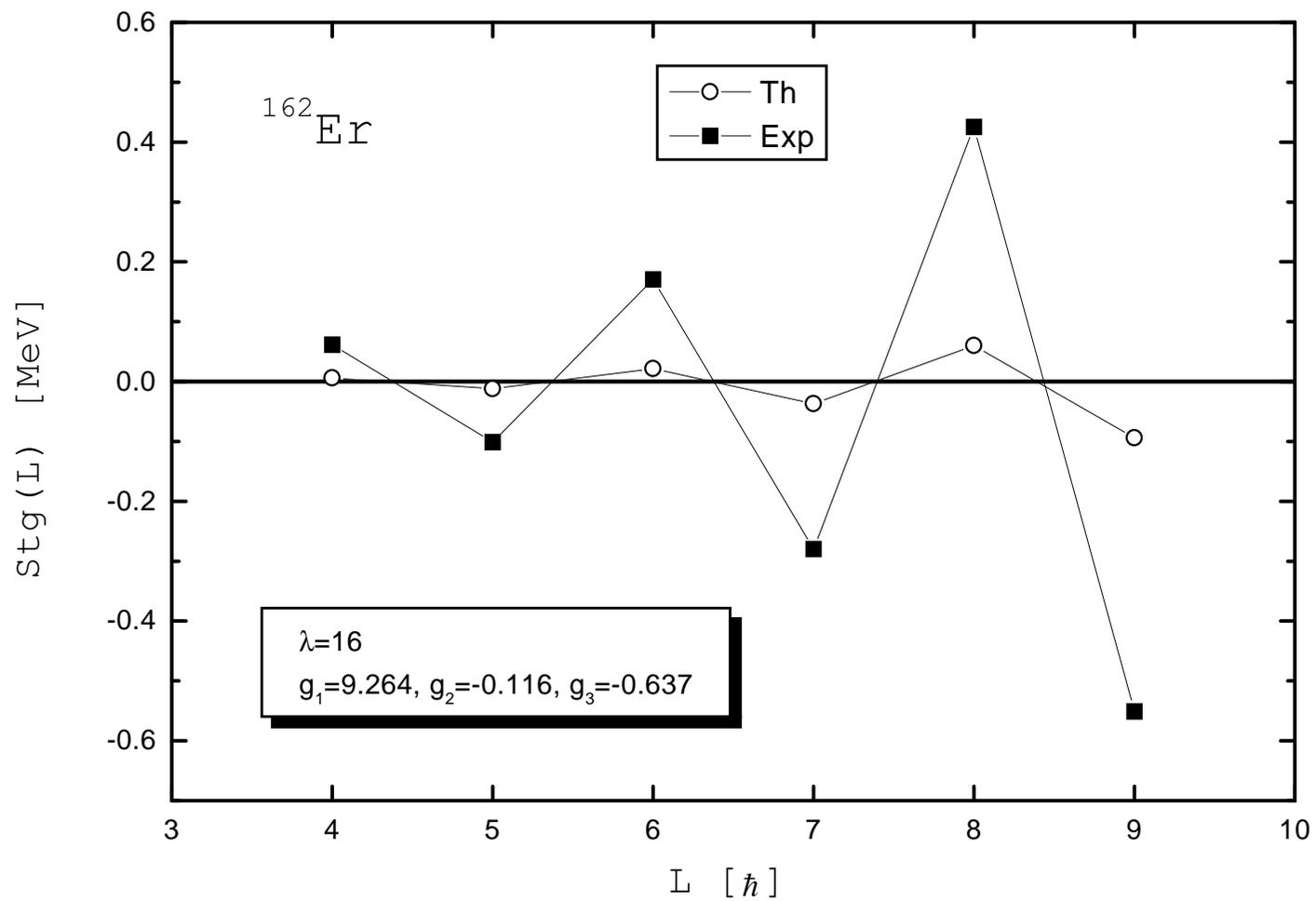
{\bf Figure 5.} The same as Fig. 1, but for $^{162}$Er.
\medskip

\noindent
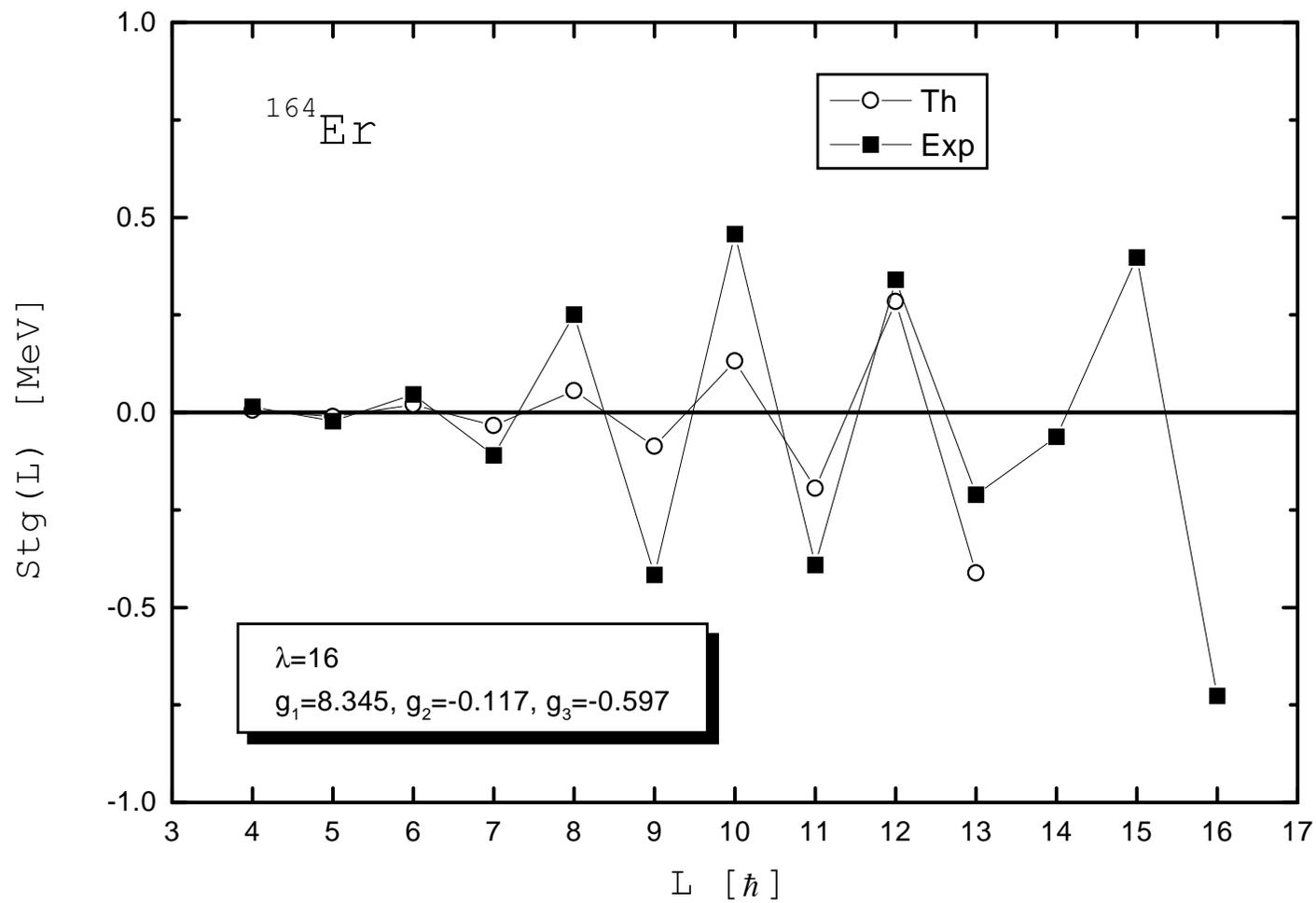
{\bf Figure 6.} The same as Fig. 1, but for $^{164}$Er.
\medskip

\noindent
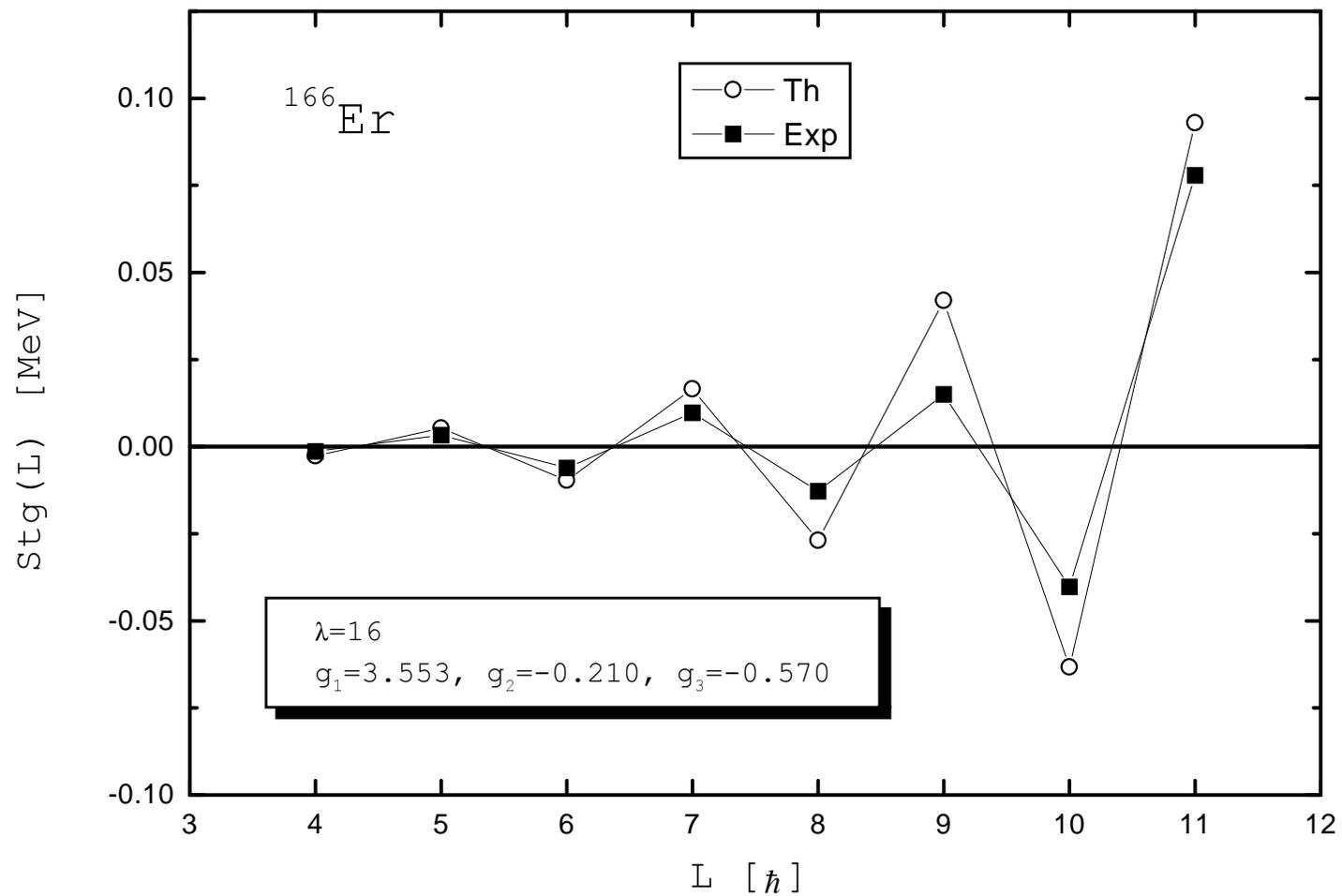
{\bf Figure 7.} The same as Fig. 1, but for $^{166}$Er.
\medskip

\noindent
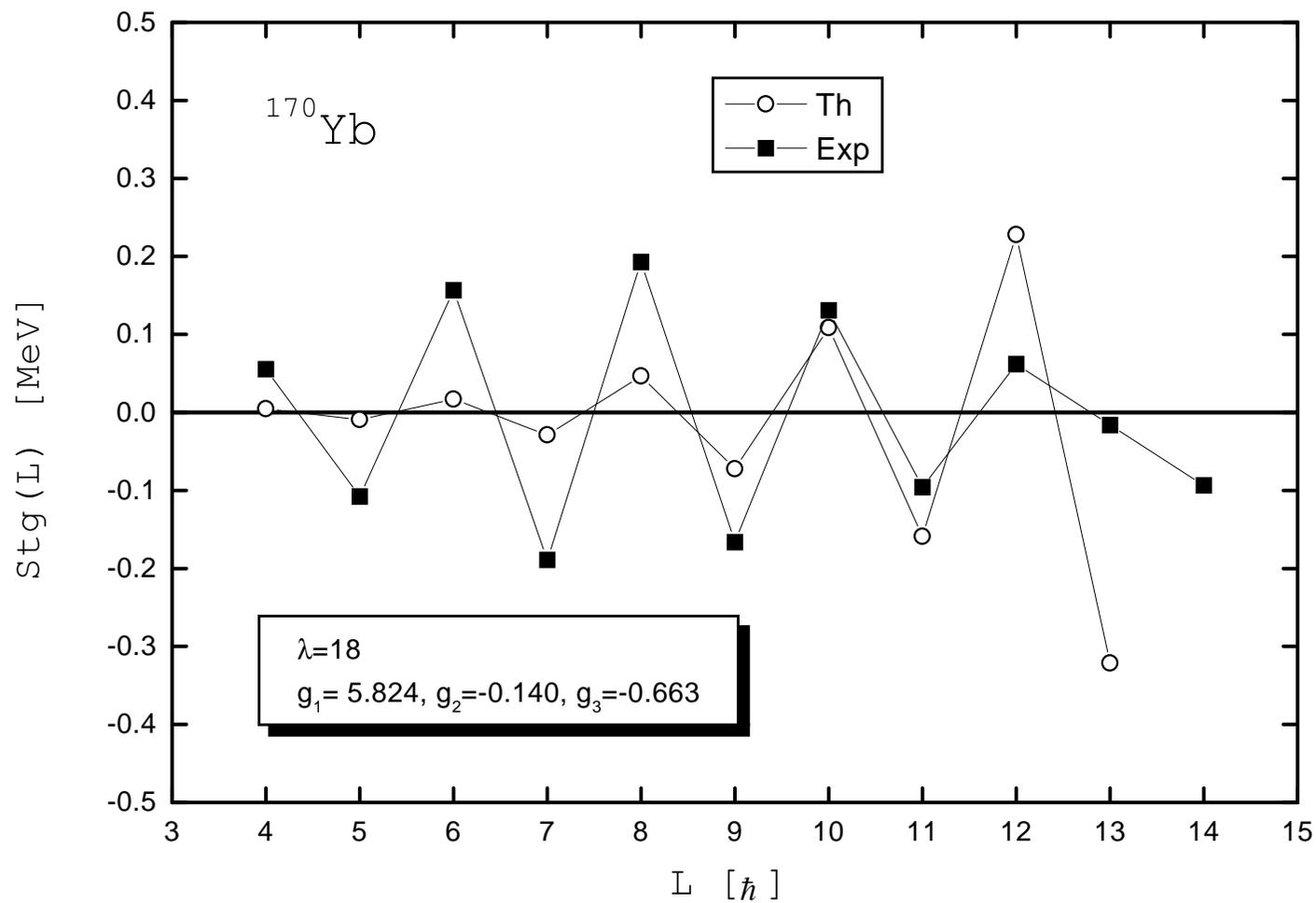
{\bf Figure 8.} The same as Fig. 1, but for $^{170}$Yb.
The experimental data are taken from \cite{Arc98}.
\medskip

\noindent
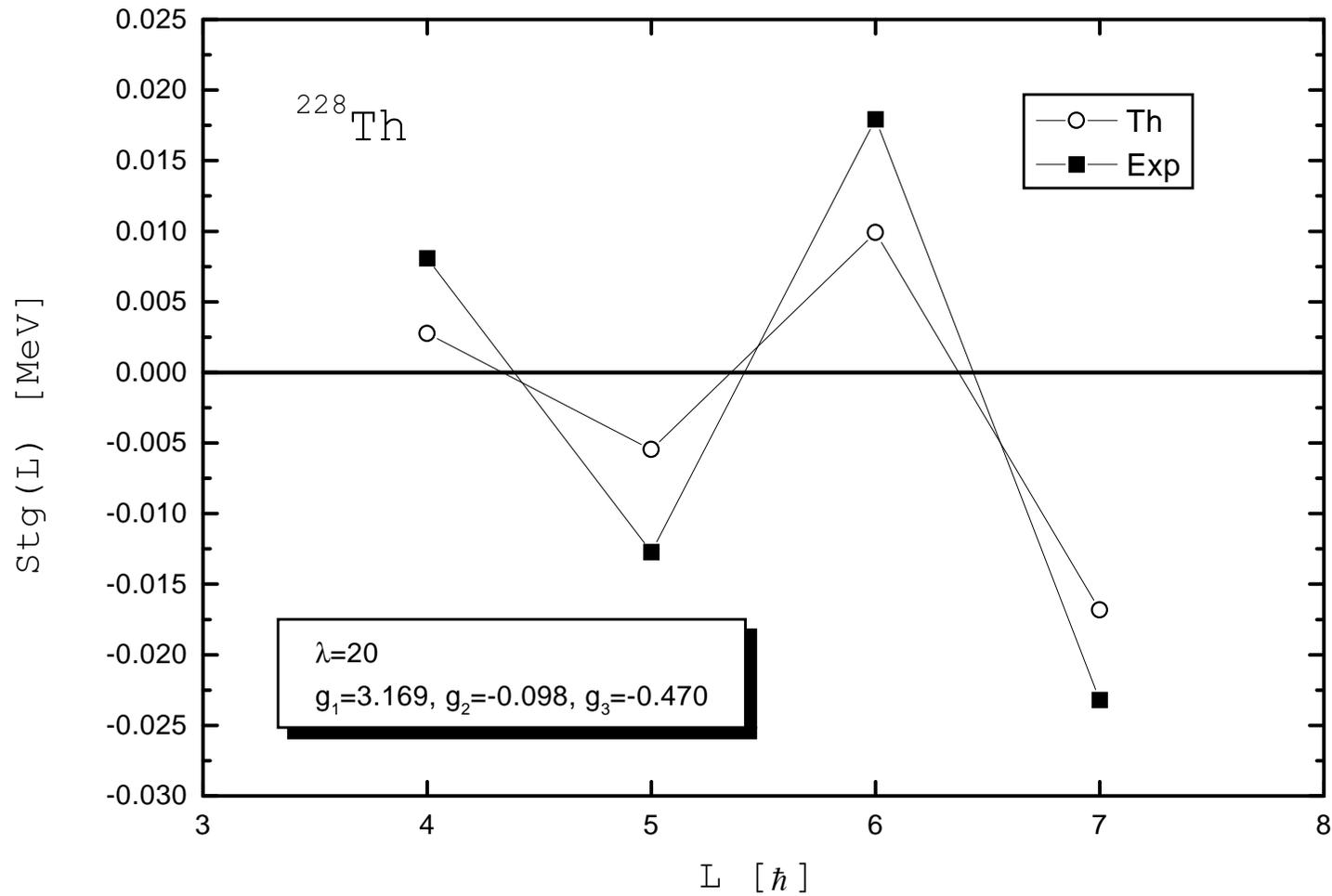
{\bf Figure 9.} The same as Fig. 1, but for $^{228}$Th.
The experimental data are taken from \cite{Web98}.
\medskip

\noindent
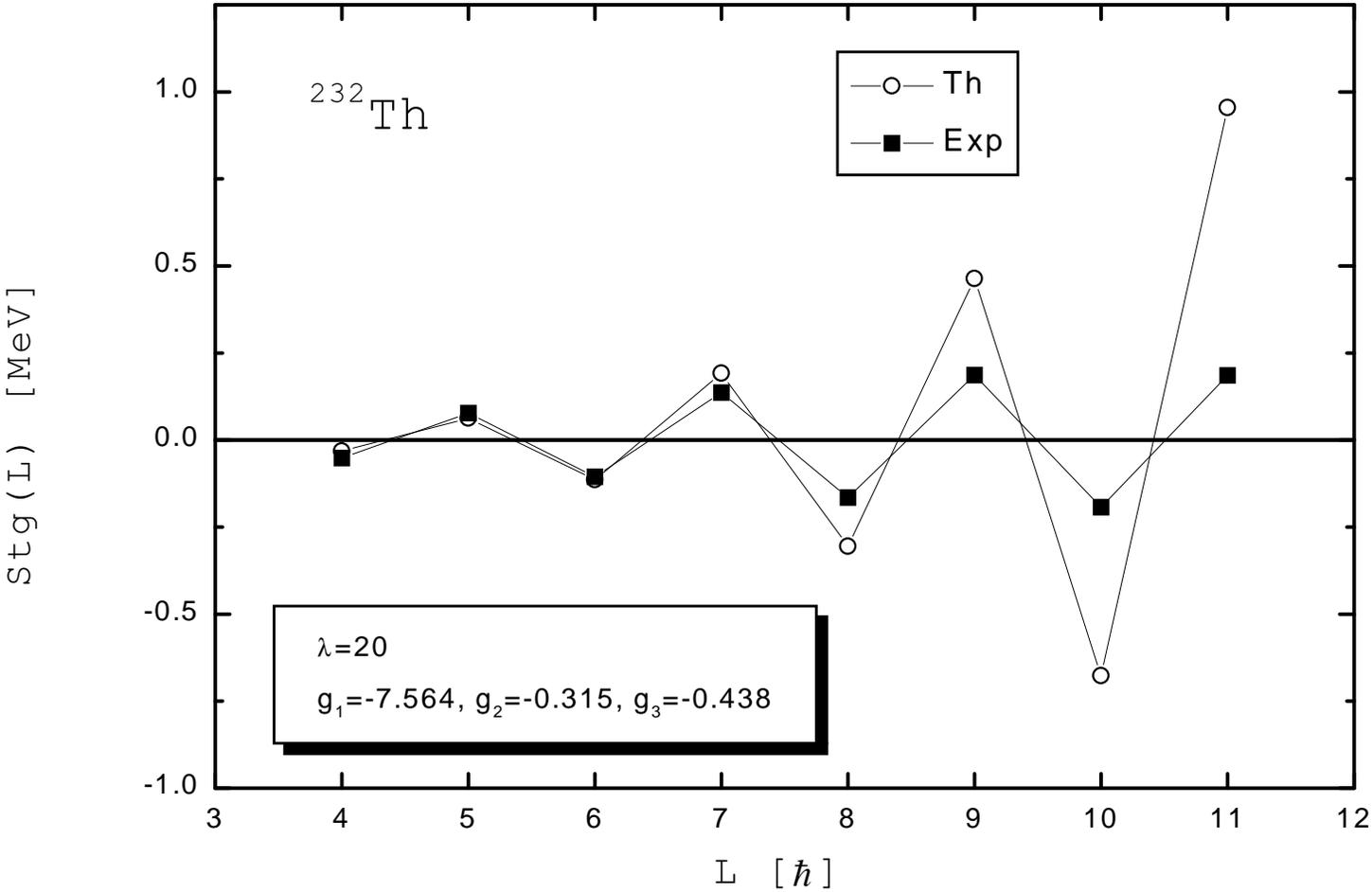
{\bf Figure 10.} The same as Fig. 1, but for $^{232}$Th.
The experimental data are taken from \cite{Sakai}.
\medskip

\noindent
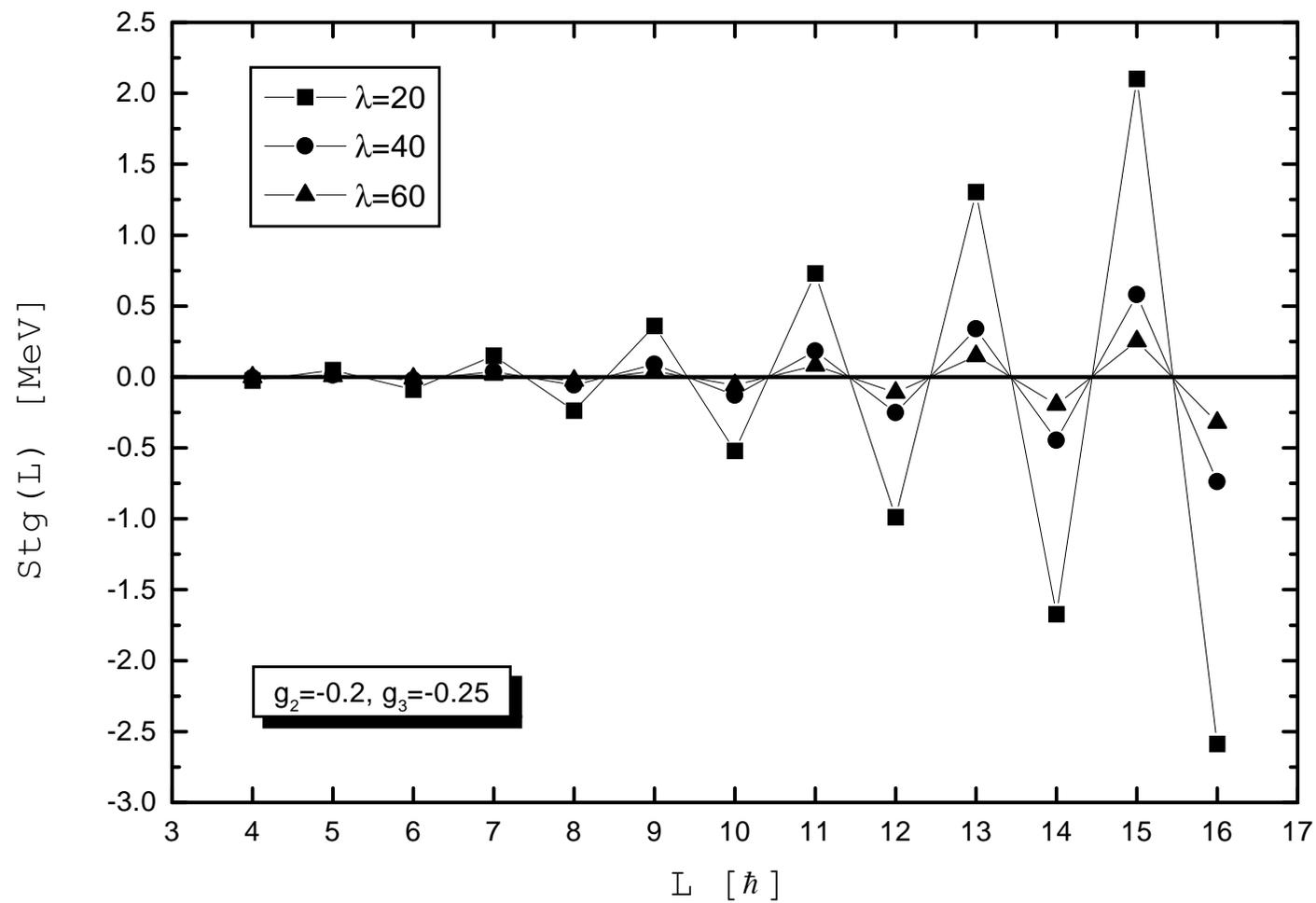
{\bf Figure 11.} The theoretical values of the quantity
$Stg(L)$, Eq.~(\ref{VBMstag}), obtained for three different
values of the SU(3) quantum number $\l$ ($\l =20$, $\l =40$ and $\l
=60$) and fixed (overall) values of the model parameters
($g_{2}=-0.2$ and $g_{3}=-0.25$) are plotted as functions of the
angular momentum.


\newpage

\begin{thebibliography}{xx}

\bibitem{BM75} A.  Bohr and B.  R.  Mottelson, {\it Nuclear Stucture}
vol. II (Benjamin, New York, 1975).

\bibitem{RS80} P.  Ring and P.  Shuck, {\em The Nuclear Many-Body
Problem} (Springer, Heidelberg, 1980).

\bibitem{Fli} S.  Flibotte {\it et al.}, Phys.  Rev.  Lett.  {\bf
71}, 4299 (1993) ; Nucl.  Phys.  {\bf A 584} 373 (1995).

\bibitem{Ced} B.  Cederwall {\it et al.}, Phys.  Rev.  Lett.  {\bf
72}, 3150 (1994).

\bibitem{Stag} D.  Bonatsos, C.  Daskaloyannis, S.  Drenska, G.
Lalazissis, N.  Minkov, P.  Raychev and R.  Roussev, Phys.  Rev.  A
{\bf 54}, R2533 (1996).

\bibitem{WZ97} C. S. Wu and Z. N. Zhou, Phys. Rev. C {\bf 56}, 1814
(1997).

\bibitem{WT97} L. A.  Wu and H. Toki, Phys. Rev. C {\bf 56}, 1821
(1997).

\bibitem{RDM97} P. Raychev, S. Drenska and J. Maruani, Phys. Rev. A
{\bf 56}, 2759 (1997).

\bibitem{Bevod} D. Bonatsos, Phys. Lett {\bf 200B}, 1 (1988).

\bibitem{Fi82} C. A. Fields, K. H. Hicks, R. A. Ristinen, F. W. N. de
Boer, P. M. Walker, J. Borggreen and L. K. Peker, Nucl. Phys. A {\bf
389}, 218 (1982).

\bibitem{Bac82} A. Backlin et al, Nucl. Phys. A {\bf 380}, 189
(1982).

\bibitem{Fi84} C. A. Fields, K. H. Hicks, R. A. Ristinen, F. W. N. de
Boer, L. K. Peker, R. J. Peterson and  P. M. Walker,  Nucl. Phys. A
{\bf 422}, 215 (1984).

\bibitem{FHP84} C.  A.  Fields, K.  H.  Hicks and R.  J.  Peterson,
Nucl. Phys.  A {\bf 431}, 473 (1984).

\bibitem{Rie87} H. J. Riezebos, M. J. A. de Voigt, C. A. Fields, X.
W. Cheng, R. J. Peterson, G. B. Hagemann and A. Stolk, Nucl. Phys. A
{\bf 465}, 1 (1987).

\bibitem{IA} F.  Iachello and A.  Arima, {\it The Interacting Boson
Model} (Cambridge University Press, Cambridge, 1987).


\bibitem{Sakai} M. Sakai, At.  Data Nucl.  Data Tables {\bf 31}, 399
(1984).

\bibitem{Sood}  P. C. Sood, D. M. Headly and R. K. Sheline, At. Data
Nucl. Data Tables {\bf 47}, 89 (1991).

\bibitem{BMI85} G. Van den Berghe, H. E. De Meyer and P. Van Isacker,
Phys.  Rev.  C {\bf 32}, 1049 (1985).

\bibitem{PVI99} P. Van Isacker, Phys. Rev. Lett. {\bf 83}, 4269
(1999).

\bibitem{p:descr} P.  P.  Raychev and R.  P.  Roussev, Sov.  J. Nucl.
Phys.  {\bf 27}, 1501 (1978).

\bibitem{a:over} S.  Alisauskas, P.  P.  Raychev and R.  P.  Roussev,
J. Phys.  G {\bf 7}, 1213 (1981).

\bibitem{p:matr} P.  P.  Raychev and R.  P.  Roussev, J.  Phys.  G
{\bf 7}, 1227 (1981).

\bibitem{MDRRB97} N.  Minkov, S.  Drenska, P.  Raychev, R.  Roussev
and D. Bonatsos, Phys.  Rev.  C {\bf 55}, 2345 (1997).

\bibitem{MDRRB98} N.  Minkov, S.  Drenska, P.  Raychev, R.  Roussev
and D. Bonatsos,  Phys.  Rev.  C {\bf 60}, 034305 (1999).

\bibitem{SZG} Y. Sun, J.-Y. Zhang and M. Guidry, Phys. Rev. Lett.
75, 3398 (1995); Phys. Rev. C 54, 2967 (1996).

\bibitem{MQ} I. N. Mikhailov and P. Quentin, Phys. Rev. Lett. 74,
3336 (1995).

\bibitem{Mag} P. Magierski, K. Burzy\'nski, E. Perli\'nska, J.
Dobaczewski and W. Nazarewicz, Phys. Rev. C 55, 1236 (1997).

\bibitem{Kota} V. K. B. Kota, Phys. Rev. C 53, 2550 (1996).

\bibitem{Liu} Y.-X. Liu, J.-G. Song, H.-Z. Sun and E.-G. Zhao, Phys.
Rev. C 56, 1370 (1997).

\bibitem{Pav} I. M. Pavlichenkov, Phys. Rev. C 55, 1275 (1997).

\bibitem{Wu} H. Toki and L.-A. Wu, Phys. Rev. Lett. 79, 2006 (1997);
L.-A. Wu and H. Toki, Phys. Rev. C 56, 1821 (1997).

\bibitem{HM} I. Hamamoto and B. Mottelson, Phys. Lett. B 333, 294
(1994).

\bibitem{Macc} A. O. Macchiavelli, B. Cederwall, R. M. Clark, M. A.
Deleplanque, R. M. Diamond, P. Fallon, I. Y. Lee, F. S. Stephens and
S. Asztalos, Phys. Rev. C 51, R1 (1995).

\bibitem{PavFli} I. M. Pavlichenkov and S. Flibotte, Phys. Rev. C
51, R460 (1995).

\bibitem{Doenau} F. D\"onau, S. Frauendorf and J. Meng, Phys. Lett. B
387, 667 (1996).

\bibitem{Luo} W. D. Luo, A. Bouguettoucha, J. Dobaczewski, J. Dudek
and X. Li, Phys. Rev. C 52, 2989 (1995).

\bibitem{Magi} P. Magierski, Warsaw University of Technology preprint
nucl-th/9512004, to appear in Acta Physica Polonica B.

\bibitem{BieLou} L.  C.  Biedenharn and J.  D.  Louck, {\it Angular
Momentum in Quantum Physics}, Encyclopedia of Mathematics and its
Applications {\bf 8} (Addison Wesley, Reading, 1981).

\bibitem{bm:bas} V.  Bargmann and M.  Moshinsky, Nucl.  Phys.  {\bf
23}, 177 (1961).

\bibitem{m:bas} M.  Moshinsky, J.  Patera, R.  T.  Sharp and P.
Winternitz, Ann.  Phys.  (N.Y.) {\bf 95}, 139 (1975).

\bibitem{Arc98} D. E. Archer, M. A. Riley, T. B. Brown, D. J.
Hartley, J. D\"oring, G. D. Johnes, J. Pfohl, S. L. Tabor, J.
Simpson, Y. Sun, and J. L. Egido, Phys. Rev. C 57, 2924 (1998).

\bibitem{Web98} T. Weber, J. Gr\"oger, C. G\"unther and J. deBoer,
Eur. Phys. J. A {\bf 1}, 39 (1998).

\bibitem{166Er1} C. W. Reich and J. E. Gline, Nucl. Phys. A {\bf
159}, 181 (1970).

\bibitem{166Er2} C. W. Reich and J. E. Gline, Phys. Rev. {\bf 129},
2152 (1963).

\bibitem{NDS166} A. E. Ignatochkin, E. N. Shurshikov and Yu. F.
Jaborov, Nucl. Data Sheets {\bf 52}, 365 (1987).

\bibitem{Gilmore} R. Gilmore, {\em Lie Groups, Lie Algebras and Some
of Their Applications} (Wiley, New York, 1974).

\bibitem{CBVR86} J. Carvalho, R. Le Blanc, M. Vassanji and
D. J. Rowe, Nucl. Phys. {\bf A452}, 240 (1986).

\bibitem{CDL88} O. Casta\~{n}os, J. P. Draayer and Y. Leschber,
Z. Phys. A {\bf 329}, 33 (1988).

\bibitem{RVC89} D. J. Rowe, M. G. Vassanji and J. Carvalho,
Nucl. Phys. {\bf A504}, 76 (1989).

\bibitem{Mukerjee} M.  Mukerjee, Phys.  Lett. {\bf 251B}, 229 (1990).

\bibitem{JPD93} J.  P.  Draayer, in {\it Algebraic Approaches to
Nuclear Structure:  Interacting Boson and Fermion Models},
Contemporary Concepts in Physics VI, edited by R.  F.  Casten
(Harwood, Chur, 1993) p.  423.

\bibitem{Jon78} N. R. Johnson, D. Kline, S. W. Yates, F. S. Stephens,
L. L. Riedinger and R. M. Ronningen,  Phys. Rev. Lett. 40,
151 (1978).

\bibitem{Yat80} S. W. Yates et al., Phys. Rev. C 21, 2366 (1980).

\bibitem{Bon85} D. Bonatsos, Phys. Rev. C 31, 2256 (1985).

\end{thebibliography}
\end{document}